\documentclass[a4paper,preprint,showpacs,amssymb,superscriptaddress]{revtex4}
\usepackage{graphicx}
\newcommand{\nd}{\noindent}
\newcommand{\be}{\begin{equation}}
\newcommand{\ee}{\end{equation}}
\newcommand{\ben}{\begin{eqnarray}}
\newcommand{\een}{\end{eqnarray}}

\newcommand{\tr}{{\mathrm{Tr}}}
\begin{document}
\draft
\title{Wootters' distance revisited: a new  distinguishability
criterium}
\author{A. Majtey}
\address{Facultad de Matem\'atica, Astronom\'\i a y F\'\i sica \\
Universidad Nacional de C\'ordoba \\ Ciudad Universitaria, 5000
C\'ordoba, Argentina}
\author{P.W. Lamberti}
\address{Facultad de Matem\'atica, Astronom\'\i a y F\'\i sica \\
Universidad Nacional de C\'ordoba \\ Ciudad Universitaria, 5000
C\'ordoba, Argentina}
\author{M.T. Martin}
\address{Instituto de F\'isica (IFLP), Universidad Nacional de La Plata
and CONICET, C.C. 727, La Plata 1900, Argentina}
\author{ A. Plastino}
\address{Instituto de F\'isica (IFLP),
Universidad Nacional de La Plata and CONICET, C.C. 727, La Plata
1900, Argentina}
\date{\today}

\begin{abstract}
The notion of distinguishability between quantum states has shown
to be fundamental in the frame of quantum information theory. In
this paper we present a new distinguishability criterium by using
a information theoretic quantity: the Jensen-Shannon divergence
(JSD). This quantity has several interesting properties, both from
a conceptual and a formal point of view. Previous to define this
distinguishability criterium, we review some of the most
frequently used distances defined over quantum mechanics' Hilbert
space. In this point our main claim is that the JSD can be taken
as a unifying  distance between quantum states.
\end{abstract}


\maketitle

\section{Introduction}

The problem of measurement is an issue of central importance in
quantum theory, that, since the pioneering days of the twenties
has given rise to controversies \cite{Wheeler}. Many of the most
astonishing results of quantum mechanics are related to the
particular properties of the measurement processes. In recent
years, the unique character of quantum measurement has led to a
new field of research:  quantum information technology
\cite{cryp}. From a formal point of view, a measurement in quantum
theory is described by means of an Hermitian operator. If the
eigenstates of this operator are $|\phi_k\rangle$ and the state of
the system to be measured is $|\Psi\rangle = \sum c_k
|\phi_k\rangle$, then, according to the axioms of the quantum
theory, the result of the measurement will, with probability
$|c_k|^2$, be the corresponding eigenvalue $a_k$, represented
physically by an appropriate state of the measuring device
$\cal{A}$.

A close related theme is that of the distinguishability between
states, that is, just how can we discern between two states
$|\Psi^{(1)}\rangle$ and $|\Psi^{(2)}\rangle$ of a given physical
system by using the measuring device $\cal{A}$. In a seminal
paper,
 Wootters investigated this problem and introduced a
``distinguishability-distance'' between pure states in the
associated Hilbert space \cite{Wootters}. Braunstein and Caves
extended this distance to density operators for mixed states
\cite{Caves}. Wootters distinguishability-criterium can be
established, within the framework of probability theory
(independently of any quantum interpretation), in the following
way \cite{Wootters}: two probability distributions,  say,
$p^{(1)}=(p_1,p_2,\ldots,p_N)$ and $p^{(2)}=(q_1,q_2,\ldots,q_N)$
are distinguishable after $L$ trials ($L \rightarrow \infty$) if
and only if the condition
\begin{equation} \label{fund}
\frac{\sqrt{L}}{2} \{\sum_{i=1}^N \frac{(\delta
p_i)^2}{p_i}\}^{1/2} >1
\end{equation}
with $\delta p_i = p_i -q_i$, is satisfied. This
distinguishability-criterium involves a distance defined over the
space of probability distributions \be \label{W1}
ds(p^{(1)},p^{(2)}) = \frac{1}{2}\sqrt{\sum_i \frac{(\delta
p_i)^2}{p_i}}. \ee Statisticians call to the square of this form
the $\chi^2$ distance. Wootters maps this distance into the
associated Hilbert space and establishes a correspondence with the
usual notion of distance between states in  Hilbert's space.

In addition to its relevance with regards to the
distinguishability issue, the concept of distance between
different states in a Hilbert space plays an important role in a
diversity of circumstances
\begin{itemize} \item the study of the geometric properties of the
quantum evolution sub-manifold \cite{Anandan,Abe}, \item  in
discussing squeezed coherent states or generalized coherent spin
states \cite{Kwek}, \item  in ascertaining the quality of
approximate treatments \cite{Casas1}. \end{itemize}

It has recently been recognized that the concept of
distinguishability is basic to manipulate information in the sense
that being able to discern between different physical states of a
given system allows one to determinate just how much information
can be encoded into that system, so that  the notion of
distinguishability builds a bridge between quantum theory and
information theory \cite{Vedral}.

In this work we will try to strengthen this connection by
investigating the relation between  Wootters' distance and a
suitable metric for the probability-distributions' space that is
used  in information theory: the Jensen-Shannon divergence (JSD).
Recently, the JSD has been exhaustively studied in different
contexts \cite{Grosse}. It has many interesting interpretations,
both in the framework of information theory as in the context of
mathematical statistics. One of its basic properties is that its
square root is a true metric in the probability-distributions'
space, i.e., its square root is a distance that verifies the
triangle inequality \cite{Endres}. This fact is quite relevant,
since  metric properties are crucial for the application of many
important convergence theorems that one needs when iterative
algorithms are studied.

The purpose of this paper is twofold: \begin{enumerate} \item
first, we pursue a pedagogical objective by reviewing some
distances and metrics commonly used in quantum theory. Even though
many of the results presented here are known, they are not always
presented from an unified perspective, at least in physics
literature, \item second, we formulate a distinguishability
criterium for quantum mechanics based on the JSD.
\end{enumerate}
Finally,  some conclusions are drawn.

\section{A primer on Hilbert space distances}

Let $|\phi_1\rangle,...|\phi_N\rangle$ be the eigenstates of a
given Hermitian operator associated with the measuring instrument
$\cal{A}$. For simplicity's sake we assume that no degeneration
exists. Thus, in a given measurement $N$ possible results may
ensue. If we have prepared the system in the (normalized) state
$|\Psi^{(1)}\rangle$, each of these results can be found with
probability $|\langle\phi_i|\Psi^{(1)}\rangle|^2$. If we prepare
it, instead, in the state $|\Psi^{(2)}\rangle$, this probability
is $|\langle\phi_i|\Psi^{(2)}\rangle|^2$. Since the basis
$|\phi_i\rangle$ is complete \be \label{4}  \sum_i
|\phi_i\rangle\langle\phi_i| = I, \ee one has \be \label{5}
\sum_i|\langle\phi_i|\Psi^{(1)}\rangle|^2 =
\sum_i|\langle\phi_i|\Psi^{(2)}\rangle|^2 = 1. \ee Let us write
\begin{eqnarray} \label{6}
p_i^{(1)} &=& |\langle\phi_i|\Psi^{(1)}\rangle|^2 \nonumber \\
p_i^{(2)} &=& |\langle\phi_i|\Psi^{(2)}\rangle|^2
\end{eqnarray}
\nd An alternative way of looking at things is as follows. Let \be
\label{7}  X_N^+=\{(p_1,\ldots,p_N); 0\leq p_i\leq 1; \sum_i p_i
=1\} \ee be the set of discrete probability distributions
(generalization to  continuous ones being straightforward) and let
$\cal{S}$ be the set of normalized states in the Hilbert space
${\cal{H}}^{n+1},\;\;n+1=N$. To each states $|\Psi\rangle$ in
$\cal{S}$ (indeed to a ray $\lambda|\Psi\rangle, \; \lambda =
e^{i\varphi}$) we assign an element $\{p_i\}$ of $X_N^+$ through
the application ${\cal{F}}_{\cal{A}}$ given by: \ben \label{8}
\mathcal{F}_{\cal{A}} : \mathcal{S} \subset {\mathcal{H}}^{n+1}
&\rightarrow& X_{\rm N}^+  \cr \vert \Psi\rangle &\rightarrow&
\{p_i\} \,\, \,\, \,\, such\,\, that\,\, p_i =\vert
\langle\phi_i|\Psi\rangle \vert^2. \label{eq1} \een Obviously, the
application $\mathcal{F}_{\cal{A}}$ is consistent with expressions
(\ref{5}) and (\ref{6}).

Let $s_X(p^{(1)},p^{(2)})$ be a distance defined on the space of
probability distributions $X_N^+$, that is, an application from
$X_N^+\times X_N^+$ into $\Re$ such that is symmetric and
$s_X(p^{(1)},p^{(2)})=0$ if and only if $p^{(1)}=p^{(2)}$. One can
associate to $s_X(p^{(1)},p^{(2)})$ a distance in the space
${\cal{H}}^{n+1}$, $s_{\cal{H}}^{\cal{A}}(|\Psi^{(1)}\rangle,
|\Psi^{(2)}\rangle)$ through the application
$\mathcal{F}_{\cal{A}}$. Let us note that this distance depends
upon the measuring instrument $\cal{A}$. Our objective is to find
a representative distance of $s_X(p^{(1)},p^{(2)})$ in  Hilbert's
space independently of the basis $|\phi_k\rangle$. This will be
attained by looking for the maximum of the associated distance
$s_{\cal{H}}^{\cal{A}}$. We discuss some examples below. The
pertinent distances are given proper names (e.g., Wootters),
according to common usage.

\textit{Notation remark}: We will use the following notation:
$s_X$ denotes a distance defined over $X_N^+ $;
$s_{\cal{H}}^{\cal{A}}$ denotes the corresponding distance over
${\cal{H}}^{n+1}$ obtained from the correspondence induced by
application ${\cal{F}_{\cal{A}}}$; $S_{\cal{H}}$ denotes the
maximum of $s_{\cal{H}}^{\cal{A}}$.

\subsection{Wootters´ distance}

The Wootters distance between two probability distributions,
$p^{(1)}$ and $p^{(2)}$ is defined as \be \label{9}
s^W_X(p^{(1)},p^{(2)}) = \arccos(\sum_i \sqrt{p_i^{(1)}
p_i^{(2)}}). \ee When $p^{(1)} \rightarrow p^{(2)}$, the form
(\ref{W1}) is reobtained.

By using the correspondence (\ref{eq1}), we can write \be
\label{10} s_{\cal{H}}^{W,\cal{A}}(|\Psi^{(1)}\rangle,
|\Psi^{(2)}\rangle)= \arccos(\sum_i
|\langle\phi_i|\Psi^{(1)}\rangle||\langle\phi_i|\Psi^{(2)}\rangle|).
\ee Note that  $\arccos(x)$  decreases in $[0,1]$. Also, the
following inequality
\begin{equation}
\sum_i
\vert\langle\phi_i\vert\Psi^{(1)}\rangle\vert\vert\langle\phi_i\vert\Psi^{(2)}\rangle\vert
\geq \vert\langle\Psi^{(1)}\vert\Psi^{(2)}\rangle\vert,
\label{desig}
\end{equation}
is true for all $\{\vert\phi_i\rangle\}$. Indeed, assume
$\vert\Psi^{(1)}\rangle=\sum_k a_k \vert\phi_k\rangle$, and
$\vert\Psi^{(2)}\rangle = \sum_k b_k \vert\phi_k\rangle$. Then,
\begin{eqnarray} \label{11}
\vert\langle\Psi^{(1)} \vert\Psi^{(2)}\rangle\vert &=& \vert\sum_k
a_k b_k^* \vert
\leq \sum_k  \vert a_k b_k^*  \vert \nonumber \\
 & \leq & \sum_k
\vert\langle\phi_k\vert\Psi^{(1)}\rangle\vert\vert\langle\phi_k\vert\Psi^{(2)}\rangle\vert.
\end{eqnarray}
 Inequality (\ref{11}), together with the $\arccos-function$ decreasing
nature,
 implie that the distance
\begin{equation}
S_{\cal{H}}^W(\vert\Psi^{(1)}\rangle,
\vert\Psi^{(2)}\rangle)=\arccos(\vert\langle\Psi^{(1)}\vert\Psi^{(2)}\rangle\vert),
\label{Wootters}
\end{equation}
maximizes $s_{\cal{H}}^{W,\cal{A}}$. In this way we arrive at the
distance associated to the Wootters' one in Hilbert's space.
Geometrically, it gives the angle between the two states (rays)
$|\Psi^{(1)}\rangle$ and $|\Psi^{(2)}\rangle$.

\subsection{Hellinger' distance:}

Let $s^H_X$ be a distance in $ X_N^+$ such that its square reads
\be \label{12} (s^H_X)^2(p^{(1)},p^{(2)}) = \frac{1}{2} \sum_i
\vert\sqrt{p_i^{(1)}} - \sqrt{p_i^{(2)}}\vert^2, \ee  Its
${\cal{H}}^{n+1}-$counterpart $s^{H,A}_{\cal{H}}$ satisfies \be
\label{13} (s^{H,A}_{\cal{H}})^2 = \frac{1}{2} \sum_i
\{\vert\langle\phi_i\vert\Psi^{(1)}\rangle\vert -
\vert\langle\phi_i\vert\Psi^{(2)}\rangle\vert \}^2, \ee that can
be cast as \be \label{14} 1-\sum_i
\vert\langle\phi_i\vert\Psi^{(1)}\rangle\vert\langle\phi_i\vert\Psi^{(2)}\rangle\vert.
\ee We see that, according to the inequality  (\ref{desig}), the
distance \be \label{15} (S^H_{\cal{H}})^2(\vert\Psi^{(1)}\rangle,
\vert\Psi^{(2)}\rangle)=1 -
\vert\langle\Psi^{(1)}\vert\Psi^{(2)}\rangle\vert, \ee is the
maximum of the associated  distance $s^H_X$. It is known as
Hellinger-distance and it represents the sine of the half angle
between the two Hilbert space vectors $|\Psi^{(1)}\rangle$ and
$|\Psi^{(2)}\rangle$ \cite{Brady}.

\subsection{Bhattacharyya' distance}

Another distinguishability measure arises from Bhattacharyya
coefficients. For two probability distributions $p^{(1)}$ and
$p^{(2)}$, the Bhattacharyya coefficients are defined by
\cite{Bhatta}
\begin{equation}
B(p^{(1)},p^{(2)})= \sum_i \sqrt{p^{(1)}_i} \sqrt{p^{(2)}_i}
\end{equation}
 Out of  these coefficients we can define a distance between
probability distributions: \be \label{16}
s^B_X(p^{(1)},p^{(2)})=-\ln(B(p^{(1)},p^{(2)})). \ee \nd  Note
that the Wootters' distance can be also expressed in terms of the
coefficients $B(p^{(1)},p^{(2)})$ as $s^W_X(p^{(1)},p^{(2)}) =
\arccos(B(p^{(1)},p^{(2)}))$. It is worth  mentioning that neither
Wootters' nor the distance  (\ref{16}) are metrics because they do
not verify the triangle inequality.

The associated distance to (\ref{16}) in Hilbert's space is
\begin{equation}
\label{16bis} s_{\cal{H}}^{B,{\cal{A}}}=-\ln\sum_i
\vert\langle\phi_i\vert\Psi^{(1)}\rangle\vert\vert\langle\phi_i\vert\Psi^{(2)}\rangle\vert.
\end{equation}
Now, since the function $-\ln(x)$ decreases with $x$, on the basis
of (\ref{desig}) we gather that  \be \label{17} S_{\cal{H}}^B
(\vert\Psi^{(1)}\rangle, \vert\Psi^{(2)}\rangle)=
-\ln\vert\langle\Psi^{(1)}\vert\Psi^{(2)}\rangle\vert, \ee is the
maximum of Bhattacharyya's distance.

\nd In these examples we focused attention upon the maximums.
Also, we have been able to cast all these distances as a
function of a Riemannian Hilbert-space metric: an ``angle" between
rays, the only one that remains  invariant under the action of the
time-evolution unitary operator.

\subsection{ Fubini-Study's metric}

Let us recall that the Hilbert space ${\cal{H}}^{n+1}$  is
isomorphic to the $n$-dimensional complex projective space
${\cal{P}}^n$, that is, the quotient space \be \label{3}
{\cal{P}}^n = (C^{n+1}-\{0\})/\sim. \ee  with $\sim$ the
equivalence relation given by \be  \label{2}
|\psi\rangle\sim|\phi\rangle \;\; iff
\;\;\exists\,\lambda\,\varepsilon \,C-{0} \; such\;\;that\;\;
|\psi\rangle=\lambda\,|\phi\rangle. \ee In this example we start
with a ${\cal{H}}^{n+1}-$ distance and construct one in $X_N^+$
(previously we proceeded in reverse fashion). In ${\cal{P}}^n$ one
defines the Fubini-Study metric $\theta_{FS}$ according to
\begin{equation}
\cos^2(\frac{\theta_{FS}}{2})\equiv
\frac{\langle\psi\vert\eta\rangle\langle\eta\vert\psi\rangle}
{\langle\psi\vert\psi\rangle\langle\eta\vert\eta\rangle},
\label{fs}
\end{equation}

For $\vert\psi\rangle\sim\vert\phi\rangle$, one has
$\theta_{FS}=0$. Maximum separation between two states is attained
for $\theta_{FS} = \pi$. Let i) ${\cal{S}}\subset {\cal{P}}^n $ be
the set of normalized states in ${\cal{P}}^n$ while ii)
$\vert\psi\rangle$ and  $\vert\psi\rangle+ \vert d\psi\rangle$ are
two very close states in $\cal{S}.$ Normalization implies
\begin{equation} \label{18}
2 Re (\langle\psi \vert d\psi\rangle) = -\langle d\psi \vert
d\psi\rangle.
\end{equation}
>From (\ref{fs}), by putting $\vert\eta\rangle=\vert\psi\rangle+
\vert d\psi\rangle$, we can evaluate the Fubini-Study distance
between two infinitely close states: \be \label{19}
\cos^2(\frac{d\theta_{FS}}{2})
\simeq(1-\frac{1}{2!}(\frac{d\theta_{FS}^2}{2})+....)^2\simeq1-\frac{(d\theta_{FS}^2)}{4},
\ee so that
\begin{equation} \label{20}
d\theta_{FS}^2=4(\langle d\psi \vert d\psi\rangle-\vert\langle\psi
\vert d\psi\rangle\vert^2).
\end{equation}
If $ \vert d\psi_\bot\rangle \equiv
 \vert d\psi\rangle-\vert\psi\rangle\langle\psi \vert d\psi\rangle$ is
 the orthogonal projection onto $\vert \psi\rangle$ of $ \vert
d\psi\rangle$, the Fubini-Study metrics acquires the aspect
\cite{Caves} \be \label{21} d\theta_{FS}^2 = 4\langle d\psi_\bot
\vert d\psi_\bot\rangle.\ee

An alternative approach to the Fubini-Study metric can be found in
reference \cite{Anandan}.

Assume now the following expansions for $\vert\psi\rangle$ and
$\vert\eta\rangle = \vert\psi\rangle+\vert d \psi\rangle$:
\begin{eqnarray} \label{29}
\vert\psi\rangle &=& \sum_i \sqrt{p_i} \;\vert\phi_i\rangle
\nonumber
\\
\vert\eta\rangle &=&\sum_i \sqrt{p_i+dp_i} \; \vert\phi_i\rangle,
\end{eqnarray}
noticing that one might add appropriate phases in both equations.
These phases, however, can be eliminated by a proper
basis-transformation (see reference \cite{Caves}). The
Fubini-Study distance between these states, up to second  order in
$dp_i$ becomes
\begin{equation}
 d\theta^2_{FS}(|\psi\rangle,|\eta\rangle) = \frac{1}{4}\sum_i
\frac{dp_i^2}{p_i}. \label{inffs}
\end{equation}
which can be thought as the corresponding Fubini-Study metric
between the distributions $\{p_i\}$ and $\{p_i+dp_i\}$ over the
space $X_N^+$.

\section{Jensen-Shannon divergence}

Information theoretic measures allow one to build up quantitative
entropic divergences between two probability distributions. A
common entropic measure is the Kullback-Leibler divergence:
\begin{equation}
s_X^K(p^{(1)},p^{(2)})= \sum_i p_i^{(1)} \ln\frac{p_i^{(1)}}{p_i^{(2)}}
\end{equation}
This distance, however, is i) not symmetric, ii) unbounded, and iii) not
always well defined. To overcome these limitations Rao and  Lin
introduced a symmetrized version of the Kullback-Leibler
divergence  the Jensen-Shannon divergence (JSD), which is defined
as \be \label{30} s_X^{JS}(p^{(1)},p^{(2)}) =
H(\frac{p^{(1)}+p^{(2)}}{2})-\frac{1}{2}H(p^{(1)})-\frac{1}{2}H(p^{(2)}),
\ee where  $H(p)=-\sum_i p_i \ln p_i$ stands for Shannon`s entropy
\cite{Rao},\cite{Lin}.

The minimum of the JSD occurs at $p^{(1)} = p^{(2)}$ and its
maximum is reached when $p^{(1)}$ and $p^{(2)}$ are two distinct
deterministic distributions. In this case $s_X^{JS}= ln 2$. As it
was mentioned previously, one of the JSD main properties is that
of being the square of a metric. A proof of this fact can be found
in reference \cite{Endres}. Alternatively, this can be proved
starting from some classical results of harmonic analysis due to
Schoenberg \cite{Schoe},\cite{Berg}.  The basic property of the
JSD that makes Schoenberg theorem applicable is that $s_X^{JS}$ is
a definite negative kernel, that is, for all finite collection of
real number $(\zeta_i)_{i \leq N}$ and for all corresponding
finite sets $(x_i)_{i \leq N}$ of points in $X_N^+$, the
implication
\begin{equation}
\sum_i^N \zeta_i =0 \Rightarrow\sum_{i,j} \zeta_i \zeta_j
s_X^{JS}(x_i,x_j)\leq 0
\end{equation}
is valid \cite{Tops1}.

Another consequence of Schoenberg's theorems is that the metric
space $(X_N^+, \sqrt{s_X^{JS}})$ can be isometrically mapped into
a subset of a Hilbert space. This result establishes  a connection
between information theory and differential geometry \cite{Tops2},
which could have interesting consequences in the realm of quantum
information theory.

Consider once again the states $\vert\psi\rangle$ and
$\vert\eta\rangle$ given by (\ref{29}) in order to evaluate the
JSD between the concomitant probability distributions
$p^{(1)}(|\psi\rangle),\,\,p^{(2)}(|\eta\rangle)$. By doing so  we
are evaluating the associated distance in Hilbert's space
$s_{\cal{H}}^{JS,{\cal{A}}}$ between the states $\vert\psi\rangle$
and $\vert\eta\rangle$.  Expanding the pertinent JSD in
$dp_i$-terms, one easily ascertains that the first non-vanishing
contributions are the quadratic ones
\begin{equation} \label{31}
ds_{\cal{H}}^{JS,{\cal{A}}}(\vert\psi\rangle, \vert\eta\rangle)  =
\frac{1}{8}\sum_i \frac{dp_i^2}{p_i},
\end{equation}
which coincides with (a half of) the Fubini-Study (\ref{inffs}) instance up to
this order in $dp_i$. Up to same order a similar relation exits
between the JSD and both the Wootters' and the Bhattacharyya'
distances, that is
\begin{equation}
ds_{\cal{H}}^{JS, {\cal{A}}}= \frac{1}{2}(dS_{\cal{H}}^{W,{\cal{A}}})^2 \label{scnorder}
=\frac{1}{2}(dS_{\cal{H}}^{B,{\cal{A}}})^2,
\end{equation}
which can be easily checked by inspection. Incidentally, it is
worth mentioning that, when we have a continuous probability distribution
$p(x)$, the JSD between $p(x)$ and its shifted version
$p(x+\delta)$ is related to the Fisher information measure $I$
through the expression
\begin{equation}
s_X^{JS}(p(x),p(x+\delta)) \simeq \frac{\delta}{2}
\sqrt{\frac{I}{2}}
\end{equation}
with
\begin{equation}
I[p(x)]=\int \frac{[\frac{dp(x)}{dx}]^2}{p(x)} dx
\end{equation}

\begin{figure}
\begin{center}
\includegraphics[scale = 0.5]{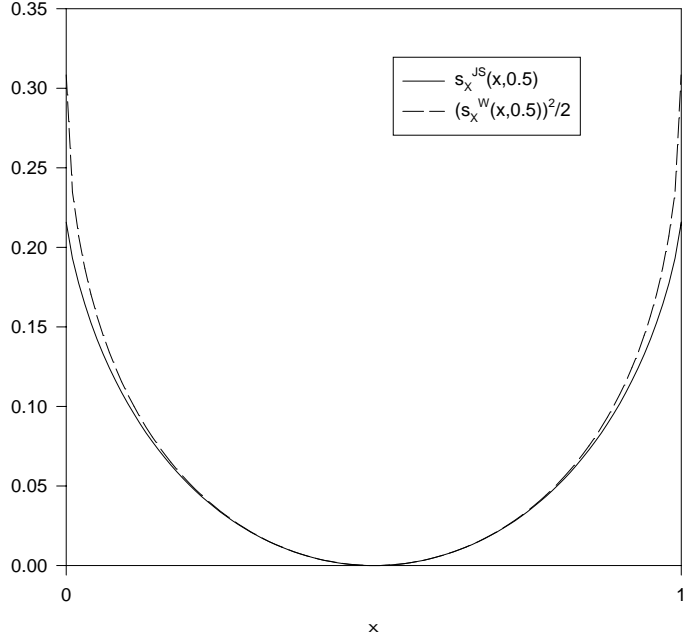}
\end{center}
\caption{Plots of $s_X^{JS}$ and $\frac{(s_X^{W})^2}{2}$. See text
for details}  \label{figure0}
\end{figure}

Equations (\ref{scnorder}) have been established up to second
order in $dp_i$. Let us proceed to higher orders. To do this let
us consider a binary system (a generalization to a system with a
greater number of states  is straightforward). Let $p^{(1)} =
(p,q)$ and $p^{(2)} = (p+dp, q-dp)$ with $p+q=1$ two neighboring
probability distributions and evaluate the pertinent JSD up to
order $dp^4$. We get
\begin{eqnarray} \label{32}
ds_X^{JS} =
-\frac{1}{8}\frac{1}{(p-1)p}dp^2+\frac{1}{16}\frac{2p-1}{p^2(p-1)^2}dp^3
- \frac{7}{192}\frac{3p^2-3p+1}{p^3(p-1)^3}dp^4+o(dp^5).
\end{eqnarray}
In turn, the corresponding Wootters' distance squared, up to the
same order is
\begin{eqnarray} \label{33}
\frac{1}{2}(ds_X^W)^2=-\frac{1}{8}\frac{1}{(p-1)p}dp^2+\frac{1}{16}\frac{2p-1}{p^2(p-1)^2}dp^3-
\frac{1}{384}\frac{44p^2-44p+15}{p^3(p-1)^3}dp^4+o(dp^5).
\end{eqnarray}
We detect coincidence between (\ref{32}) and (\ref{33}) up to
order $dp^3$. The fourth order difference equals $\frac{1}{192}$.
In other words, the relation
\begin{equation}
ds_X^{JS}= \frac{1}{2}(ds_X^W)^2 \label{frthorder}
\end{equation}
can be established up to third order in $dp$. Figure 1 shows how
$s_X^{JS}$ and $\frac{(s_X^{W})^2}{2}$ approache one to each other
for $p^{(1)} \approx p^{(2)}$. We took $p^{(1)} = (a,1-a)$ and
$p^{(2)} = (b,1-b)$ and evaluated the corresponding distances as a
function of $b$ by fixing $a=0.5$.

Going back to Wootters' distinguishability criterium (\ref{fund}),
with equation (\ref{frthorder}) in mind, we are in a position to
enunciate an alternative criterium: two probability distributions
$P^{(1)}$ and $P^{(2)}$ are distinguishable after $L$ trials
($L\rightarrow \infty$) if and only if
\begin{equation} \label{new}
(s_X^{JS}(P^{(1)},P^{(2)}))^{1/2} > \frac{1}{\sqrt{2L}}
\end{equation}

There exist formal arguments in favor of this last statement,
namely i) $(s_X^{JS})^{1/2}$ is a true metric for the space
$X_N^+$ and ii) this criterium is established in terms of an
information theoretic quantity, the JSD. Obviously inequality
(\ref{new}) is equivalent to inequality (\ref{fund}) for two
distributions ``close" enough.

In the context of section II the following question emerges: what
metric is the representative of $s_X^{JS}$ in  Hilbert's space
${\cal{H}}^{n+1}$? Equivalently: what is the maximum of the
metric $s_{\cal{H}}^{JS,{\cal{A}}}$? In this case it is difficult
(or impossible) to obtain an analytical expression for both
metrics, $s_{\cal{H}}^{JS,{\cal{A}}}$ and its upper bound
$S_{\cal{H}}^{JS}$. Anyway, it is possible to deduce an upper
bound for $s_{\cal{H}}^{JS}$. Let us consider a Hilbert space of
dimension 2D and let $\vert\Psi^{(1)}\rangle$ and
$\vert\Psi^{(2)}\rangle$ be two arbitrary, normalized states (the
extension to a greater number of dimensions is straightforward).
We set $\vert\langle\Psi^{(1)}\vert\Psi^{(2)}\rangle\vert=
\cos\varphi$ for $\varphi\varepsilon[0,\pi/2]$, that is, $\varphi$
is the Wootters distance between $\vert\Psi^{(1)}\rangle$ and
$\vert\Psi^{(2)}\rangle$.

Let $\{\vert\phi_i\rangle\}_{i=1}^2$ be an orthonormal basis for
${\cal{H}}^2$. Any other  orthonormal basis
$\{\vert\tilde{\phi}_i\rangle\}_{i=1}^2$ can be related to
$\{\vert\phi_i\rangle\}$ via the  rotation
\begin{eqnarray} \label{38}
\vert\tilde{\phi}_1(\theta)\rangle &=& \frac{e^{i\theta}}{\surd 2}
\vert\phi_1\rangle + \frac{e^{-i\theta}}{\surd 2}
\vert\phi_2\rangle\
\nonumber \\
\vert\tilde{\phi}_2(\theta)\rangle &=& -\frac{e^{i\theta}}{\surd
2} \vert\phi_1\rangle + \frac{e^{-i\theta}}{\surd 2}
\vert\phi_2\rangle\,
\end{eqnarray}
with $\theta\varepsilon[0,2\pi]$. We set $p_i^{(j)} \equiv
\vert\langle\phi_i\vert\Psi^{(j)}\rangle\vert^2$ and
$\tilde{p}_i^{(j)}(\theta) \equiv
\vert\langle\tilde{\phi}_i(\theta)\vert\Psi^{(j)}\rangle\vert^2$.
Also, $\langle\phi_i\vert\Psi^{(j)}\rangle = \sqrt{p_i^{(j)}}
e^{i\alpha_i^{(j)}}$ (via application of (\ref{8})). A little
algebra then leads to
\begin{equation} \label{39}
\tilde{p}_1^{(1)}(\theta)=\frac{p_1^{(1)}+p_2^{(1)}}{2} +
\sqrt{p_1^{(1)}p_2^{(1)}} \cos(2\theta +
\alpha_2^{(1)}-\alpha_1^{(1)}),
\end{equation}
and
\begin{equation}  \label{40}
\tilde{p}_1^{(2)}(\theta)=\frac{p_1^{(2)}+p_2^{(2)}}{2} +
\sqrt{p_1^{(2)}p_2^{(2)}} \cos(2\theta +
\alpha_1^{(2)}-\alpha_2^{(2)}).
\end{equation}
with $\alpha_i^{(j)}$ are real numbers. Moreover,
$\tilde{p}_2^{(1)}= 1 - \tilde{p}_1^{(1)}$ y $\tilde{p}_2^{(2)}= 1
- \tilde{p}_1^{(2)}$. Without loss of generality we can take
$\vert\phi_1\rangle=\vert\Psi^{(1)}\rangle$, so that
$p_1^{(1)}=1$, $\alpha_1^{(1)}=0$, $p_2^{(1)}=0$,$\sqrt{p_1^{(2)}}
= \cos\varphi$ y $\sqrt{p_2^{(2)}}=\sin\varphi$. Thus, we can
compute
$\sqrt{2s_{\cal{H}}^{JS,\tilde{\phi}}(\tilde{p}^{(2)},\tilde{p}^{(1)}})$
as a function of $\theta$. Figure 2 plots such a  function for
different $\varphi-$values. Figure 3 depicts a 3D-plot of
$\sqrt{2s_{\cal{H}}^{JS,\tilde{\phi}}}$ as a function of $\theta$
and $\varphi$. In both cases we put $\alpha_2^{(1)}=
\alpha_1^{(2)}= \alpha_2^{(2)}=0$.

\begin{figure}
\begin{center}
\includegraphics[scale = 0.5]{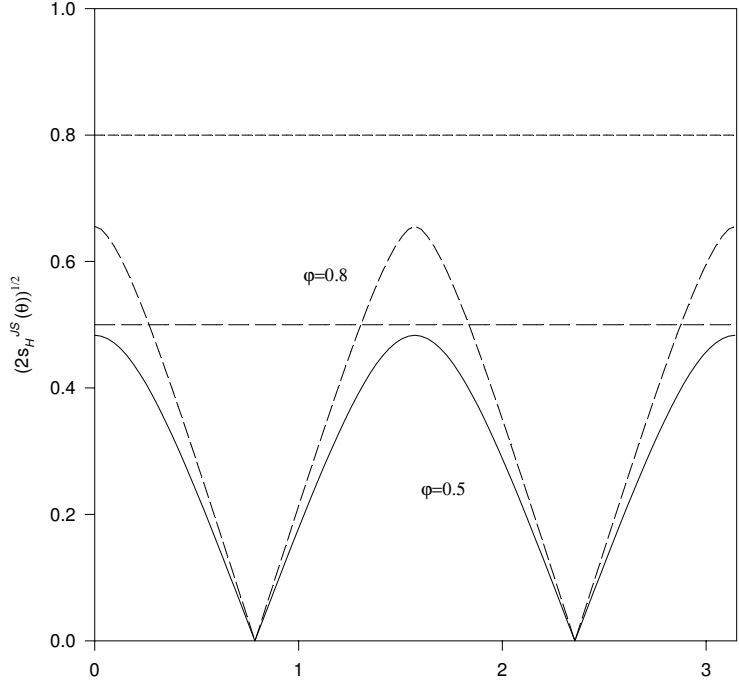}
\end{center}
\caption{\small{$\sqrt{2s_{\cal{H}}^{JS,\tilde{\phi}}(\tilde{p}^{(2)},\tilde{p}^{(1)}})$
as a function of $\theta$  for $\varphi=0.5$ and $\varphi=0.8$}}
\label{figure1}
\end{figure}

Out of  these figures we  conclude that Wootters' distance
($\varphi$) is an upper bound to
$\sqrt{2s_{\cal{H}}^{JS,\tilde{\phi}}(\tilde{p}^{(2)},\tilde{p}^{(1)}})$.
For $\varphi\rightarrow 0$, both quantities tend to coincide. In
other words, we can state the inequalities
\begin{equation} \label{fin}
S_{\cal{H}}^W(|\Psi^{(1)}\rangle,|\Psi^{(2)}\rangle) \geq
s_{\cal{H}}^{W,{\cal{A}}} (|\Psi^{(1)}\rangle,|\Psi^{(2)}\rangle)
\geq
\sqrt{2s_{\cal{H}}^{JS,{\cal{A}}}(|\Psi^{(1)}\rangle,|\Psi^{(2)}\rangle)}
\end{equation}
for any measure device $\cal{A}$.

Inequalities (\ref{fin}) allow us to conclude that
$S_{\cal{H}}^W(|\Psi^{(1)}\rangle,|\Psi^{(2)}\rangle)$
``represents'' (as the maximum, that is as the lowest upper bound)
to $\sqrt{2 s_{\cal{H}}^{JS,{\cal {A}}}}$ in the Hilbert space.
Furthermore two states distinguishable under the ``Jensen-Shannon
criterium'' are obviously distinguishable under the Wootters'
ones.

\begin{figure}
\begin{center}
\includegraphics[scale = 0.5]{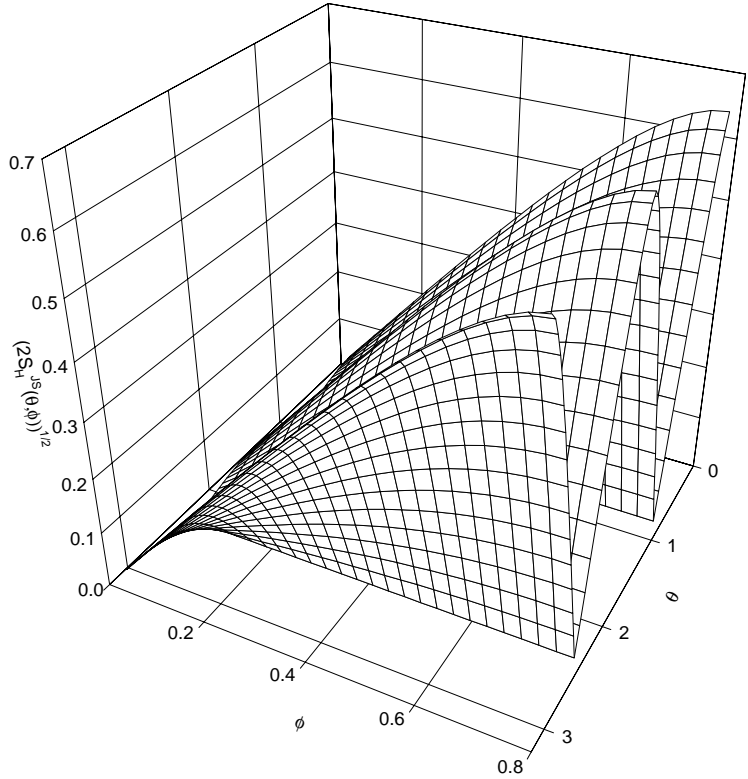}
\end{center}
\caption{\small{$3D$ Plot of
$\sqrt{2s_{\cal{H}}^{JS,\tilde{\phi}}(\tilde{p}^{(2)},\tilde{p}^{(1)}})$
as a function of $\theta$ and $\varphi$. One clearly appreciates
the bound in the plane $z=\varphi$.}} \label{figure2}
\end{figure}

\section{Conclusions}

We have proposed an alternative distinguishability criterium for
quantum states. This distinguishability criterium is established
in terms of an information theoretical quantity: the JSD, that
exhibits many interesting properties, such as a metric character
and its boundedness. This provides for a better formal context. In
some sense we feel that the JSD divergence could be taken as a
unified measure of distinguishability in the framework of quantum
information theory.

In the present work we focused on the case of pure states. An
extension to mixed states can be easily attained. In fact, by
replacing in eq.(\ref{30}) the Shannon entropy  by the von Neumann
entropy, $H_N(\rho) = - \tr(\rho \ln \rho)$, we can evaluate the
JSD between two states described by the density operators $\rho_1$
and $\rho_2$:
\begin{equation}
S_{\cal{H}}^{JS}(\rho_1,\rho_2) =
H_N(\frac{\rho_1+\rho_2}{2})-\frac{1}{2} H_N(\rho_1) - \frac{1}{2}
H_N(\rho_2) \label{mixed}
\end{equation}
Remarkably, this quantity is always well defined unlike the
corresponding Kullback-Leibler divergence that requires that the
support of $\rho_1$ is equal to or larger than that of $\rho_2$
\cite{Lindblad}. A more detailed study of the properties of JSD
for mixed states will be presented elsewhere.

Finally it is worth to mention that the JSD can be also
interpreted in a Bayesian probabilistic sense. In fact, the JSD
gives both lower and upper bounds to Bayes' probability error.
Therefore, it deserves careful scrutiny in the light of some
alternative quantum descriptions \cite{Fuchs}.

\newpage
\begin{center}
AKNOWLEDGMENT
\end{center}

We are grateful to Secretaria de Ciencia y Tecnica de la
Universidad Nacional de C\'ordoba for financial assistance. AM is
a fellowship holder of SECYT-UNC and PWL and AP are members of
CONICET. This work was partially supported by Grant
BIO2002-04014-C03-03 from Spanish Goverment.

\end{document}